# Probing the Orbital Origin of Conductance Oscillations in Atomic Chains

*Ran Vardimon, Tamar Yelin, Marina Klionsky,*
*Soumyajit Sarkar, Ariel Biller, Leeor Kronik and Oren Tal*

AUTHOR ADDRESS

Ran Vardimon
Department of Chemical Physics, Weizmann Institute of Science, Rehovot, 76100 Israel
ran.vardimon@weizmann.ac.il

Tamar Yelin
Department of Chemical Physics, Weizmann Institute of Science, Rehovot, 76100 Israel
tamar.yelin@weizmann.ac.il

Marina Klionsky
Department of Chemical Physics, Weizmann Institute of Science, Rehovot, 76100 Israel
kliomar@gmail.com

Soumyajit Sarkar
Department of Materials and Interfaces, Weizmann Institute of Science, Rehovot, 76100 Israel
soumyajit.sarkar@weizmann.ac.il

Ariel Biller
Department of Materials and Interfaces, Weizmann Institute of Science, Rehovot, 76100 Israel
ariel.biller@weizmann.ac.il

Leeor Kronik
Department of Materials and Interfaces, Weizmann Institute of Science, Rehovot, 76100 Israel
leeor.kronik@weizmann.ac.il

Oren Tal* (Corresponding author)
Department of Chemical Physics, Weizmann Institute of Science, Rehovot, 76100 Israel
oren.tal@weizmann.ac.il

KEYWORDS
electronic transport; conduction channels; atomic chain; shot noise; break junction

ABSTRACT

We investigate periodical oscillations in the conductance of suspended Au and Pt atomic chains during elongation under mechanical stress. Analysis of conductance and shot noise measurements reveals that the oscillations are mainly related to variations in a specific conduction channel as the chain undergoes transitions between zigzag and linear atomic configurations. The calculated local electronic structure shows that the oscillations originate from varying degrees of hybridization between the atomic orbitals along the chain as a function of the zigzag angle. These variations are highly dependent on the directionally and symmetry of the relevant orbitals, in agreement with the order-of-magnitude difference between the Pt and Au oscillation amplitudes observed in experiment. Our results demonstrate that the sensitivity of conductance to structural variations can be controlled by designing atomic-scale conductors in view of the directional interactions between atomic orbitals.

TABLE OF CONTENTS GRAPHC

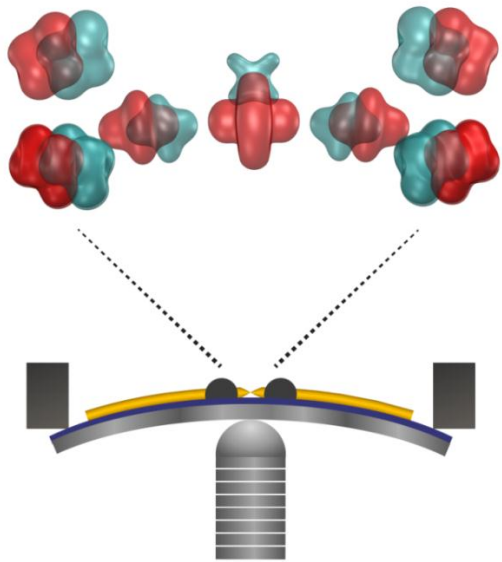

MAIN TEXT

In atomic scale conductors, electrical current is transmitted through discrete conduction channels that are dominated by the orbital structure of the conductor[1–10]. The importance of studying conductance phenomena in view of orbital structure is illustrated by the case of single atom contacts[1]. The differences in the conductance values obtained for different metal contacts were found to result from the number of atomic valence orbitals for each metal, which sets a limit to the number of conduction channels through the contact. It is therefore clear that uncovering the relation between conductance and orbital structure is essential for fundamental understanding of electronic transport through atomic scale conductors. Yet, while theoretical calculations can estimate the different conduction channels and their relation to the orbital structure[11–16], experiments are mostly limited to measuring the overall conductance.

In this Letter, we show that by obtaining the conduction channels experimentally, we can directly probe the orbital origin of conductance oscillations observed in Au and Pt atomic chains. The relative contributions of channels with different orbital character to the oscillations are inferred from an analysis of conductance and shot noise measurements. The channel resolution and observed conductance-length dependence, together with first principles calculations, allow us to associate the oscillations with the evolution of the local orbital structure as the chain undergoes transitions between zigzag and linear atomic configurations. Our findings indicate that the conductance sensitivity to structural variations results directly from the degree of spatial directionality of the orbital structure.

Atomic chains are appealing systems for the study of a large variety of phenomena, including atomic scale magnetism[17–19], nanoelectromechanics[20,21], and quantum effects in one dimension[22–25]. Chains of atoms suspended between two electrodes are particularly interesting systems for the study of electronic transport due to their relative simplicity. Such chains are observed in break junction experiments when pulling apart metallic wires of Au, Pt or Ir[26–30]. Relativistic effects taking place in these metals strengthen the interatomic bonds between the low-coordinated atoms in the atomic constriction[28,31]. As a result, atoms can be pulled out from the electrodes during the stretching process and a suspended chain of several atoms is formed between the two wire segments. First principles calculations suggest that such pulling experiments involve repeated transitions between zigzag[32] and linear atomic configurations as the chain is elongated[33–35], illustrated schematically in Fig. 1a. Such transitions are predicted to be accompanied by conductance oscillations with atomic periodicity. Here, we present experimental evidence for conductance oscillations driven by transitions between zigzag and linear atomic configurations taking place in Au and Pt atomic chains. We focus on the comparison between these metals because they possess distinct electronic transport properties resulting from the relative position of the $d$ bands with respect to the Fermi energy. We then take advantage of the correlation between structure and conductance to demonstrate the explicit relation between orbital structure and transport properties at the atomic scale.

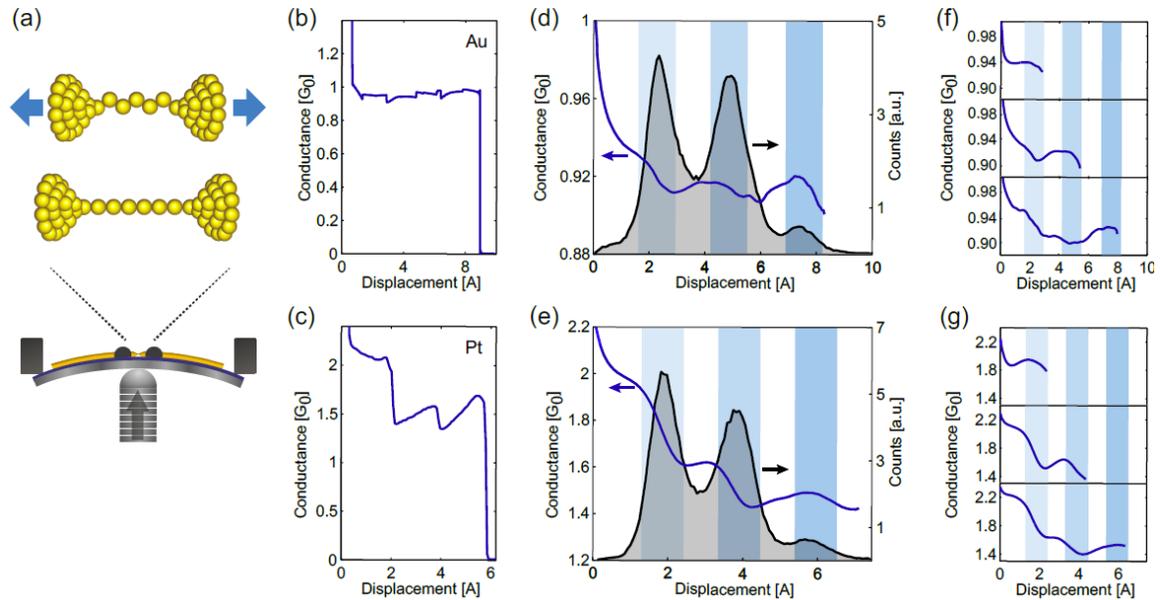

**Figure 1.** (a) Schematic illustration of transitions between zigzag and linear configurations in an atomic chain when stretched in a mechanical controllable break junction. (b,c) Examples for conductance traces showing the last plateau for Au (b) and Pt (c). (d,e) Length histogram (gray area) and average trace (blue line) for the last plateau before rupture. The trace segments used for the analysis are defined between 0.4-1.2$G_0$ for Au (d) and 1-2.5$G_0$ for Pt (e). (f,g) Average traces for 2, 3 and 4 atom long atomic chains (top, middle and bottom panel, respectively) of Au (f) and Pt (g), defined by the shaded regions enclosing the first three peaks in the respective length histogram (d,e). The presented histograms were calculated from 5,000 Au traces and 20,000 Pt traces.

We start by examining the structural and conductance properties of atomic chains that are formed when a metallic wire with a narrow constriction is pulled apart. Experiments were performed on Au and Pt wires using a break junction setup[36] at 4.2K. The conductance was recorded during each pulling sequence, after which the contact is reformed and pulled again repeatedly. Figures 1b,c show examples for conductance traces recorded for Au and Pt contacts, respectively. The traces focus on the last conductance plateau before rupture, which is attributed to a contact with a single atom in the smallest cross-section of the contact constriction[1]. The length of the plateaus can reach several Å, exceeding the typical interatomic distance. Collecting the lengths of the last plateau from thousands of pulling sequences yields the length histograms presented in Fig. 1d,e for Au and Pt, respectively. The series of peaks in the length histogram reflects the distribution of chains with different number of atoms in length. The peak positions indicate the elongations in which the tensile forces in the chain amount to a degree where it either breaks or an additional atom is pulled into the chain[20,21]. The peak separation (2.6±0.2Å for Au and 2.0±0.2Å for Pt) indicates the average distance by which the chain is stretched before another atom is added to the chain[26].

As can be observed in the individual traces (i.e. Fig. 1b,c), the conductance of the last plateau varies during elongation. In some cases, the conductance traces show oscillations with a period similar to the peak separation in the length histograms (Fig. 1d,e). In order to provide a statistical description that exposes repeated features in different traces, we calculated the average trace starting from the beginning of the last plateau. In both average traces calculated for Au and Pt (Fig. 1d,e, blue line) we find clear conductance oscillations with one-atom periodicity. Since the average trace is calculated for chains with different lengths, the oscillations in this trace could reflect contributions from the final conductance drop during the rupture process rather than actual oscillations throughout the whole elongation. In order to examine this possibility, the traces were divided into three subsets according to their total elongation. The top, middle and bottom panels of Fig. 1f(g) show the average conductance of Au(Pt) traces that break at the positions of the first, second, and third peak of the length histogram, defined by the shaded areas in Fig. 1d(e). The conductance oscillations are observed along the average traces of all subsets, despite the fact that the breaking events are now limited to the last shaded area of each trace. This behavior demonstrates that the oscillations take place during the entire course of elongation.

A straightforward explanation for the conductance oscillations is based on possible variations in the inter-atomic distance during chain elongation. In such a scenario, the conductance is expected to decrease as the chain is stretched due to larger interatomic distances which reduce the interatomic coupling[35,37] and to increase following structural relaxation once an additional atom is pulled into the chain. However, the relative positions of the observed oscillations with respect to the peaks in the length histogram (Fig. 1d,e) contradict the mentioned explanation. The conductance increases as the chain is stretched (between the peaks of the length histograms), and decreases once an additional atom enters the chain (at the positions of the peaks). The observed behavior is in very good agreement, however, with an alternative explanation based on prior calculations[33–35], in which conductance oscillations were attributed to transitions between zigzag and linear configurations. According to these calculations, the conductance increases as the relatively stable zigzag chain is stretched to a more linear configuration. Further stretching can pull a new atom into the chain, leading to partial relaxation to the less conductive zigzag configuration. We thus conclude that the experimental results support the model of repeated transitions between zigzag and linear configurations during elongation as the source for conductance oscillations.

We note that the conductance oscillations described here differ from the parity oscillations reported by Smit *et* al.[23] Parity oscillations with a double atomic periodicity were infrequently observed in our experiments for both metals, when oscillations with one-atom periodicity were suppressed (see Supporting Information[38]).

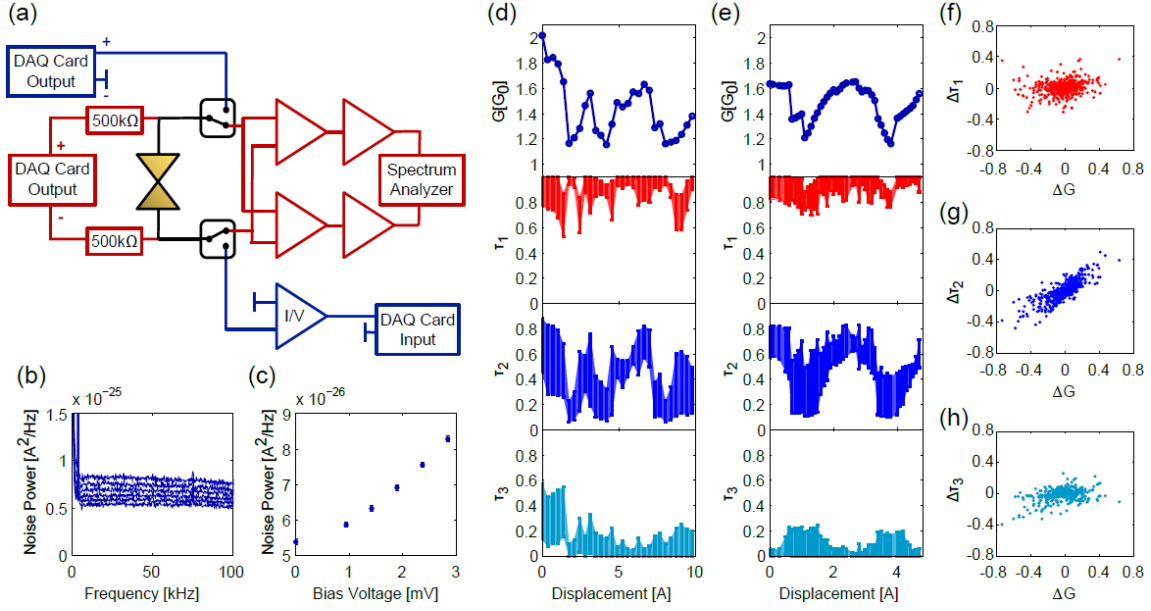

**Figure 2.** (a) Electronic setup for conductance and shot noise measurements. (b) Noise spectra recorded at different bias voltages for a Pt atomic contact with conductance of 2.32$G_0$. (c) Noise power vs. bias voltage calculated from the average noise in (b). (d,e) Pt traces showing conductance oscillations (top panel) and the experimentally resolved first three channel transmissions $\tau_1$-$\tau_3$ out of six resolved channels (lower panels). (f,g,h) Scatter plots of $\Delta\tau_n$ vs. $\Delta G$ for n=1-3 calculated from 51 channel resolved Pt traces.

To further study the underlying mechanisms for the oscillations, we obtained the distribution of conduction channels using shot noise measurements. Shot noise in quantum conductors depends on the number of channels across the conductor $N$ and their transmission probabilities $\tau_n$[39]. This relation is expressed in the Fano factor $F = \sum_{n=1}^{N} \tau_n (1 - \tau_n)/\sum_{n=1}^{N} \tau_n$, which describes the ratio between the obtained shot noise to its full Poissonian value of $2eI$ (where I is the current). The transmission probabilities also define the overall conductance $G = G_0 \sum_{n=1}^{N} \tau_n$, where $G_0 = 2e^2/h$ is the conductance quantum. Therefore, by measuring the Fano factor and the conductance it is possible to obtain two independent equations for $\tau_n$. Recently, it was shown that numerical analysis can extract the set of transmission probabilities $\{\tau_1 ... \tau_N\}$ from the two measured values[40]. This analysis can be applied to more than two channels (N>2) at the expense of the accuracy with which one can determine $\tau_n$. In general, this type of analysis can be extended to other systems that can be described by the Landauer formalism, including molecular junctions, nanotubes, nanowires, and quantum dots.

To obtain the evolution of the transmission probabilities, we measured the conductance and shot noise of Pt atomic chains during elongation. The electronic setup (Fig. 2a) is described in the Supporting Information[38]. After each elongation step, noise spectra are measured as a function of applied bias (Fig. 2b). The spectra are corrected for the RC transfer function and are averaged in a fixed frequency window (Fig. 2c). The Fano factor was calculated from the bias dependence of the noise[38]. We then employed the procedure described in reference 40 to determine the transmission probabilities.

Figures 2d,e present two typical examples for the channel resolved evolution of the conductance during the elongation of Pt atomic chains. As can be seen in the lower panels of Fig. 2d,e, the overall conductance is mainly composed of two or three conduction channels. Remarkably, throughout the elongation process the main channel maintains a transmission probability close to unity, while the transmission of the second channel clearly follows the oscillations in the conductance. Thus, the conductance oscillations are dominated by oscillations in the transmission probability of the secondary channel. We note that because $\tau_n$ are defined arbitrarily by decreasing transmission, the reference to channel identities (1st, 2nd, etc.) is based on their gradual evolution with stretching. A few points of overlap that exist between $\tau_2$ and $\tau_3$ could allow for a slightly different channel association. However, this would not change any of the presented conclusions.

In order to test whether this is a general behavior, we analyzed the conduction channels in 51 different elongation sequences. A statistical measure for the contribution of each channel to the conductance oscillations is provided by calculating the correlation between the differences in the conductance $\Delta G$ and in the channel transmissions $\Delta \tau_n$ following each elongation step. The correlation coefficient for the n$^{th}$ channel is defined as $corr(\delta G, \delta \tau_n) = \langle \delta G \delta \tau_n \rangle / \sqrt{\langle \delta G^2 \rangle \langle \delta \tau_n^2 \rangle}$ where $\delta G = \Delta G - \langle \Delta G \rangle$ and $\delta \tau_n = \Delta \tau_n - \langle \Delta \tau_n \rangle$. The scatter plots for the three main channels are shown in Fig. 2f-h. We find that the correlation coefficients for $\tau_1$, $\tau_2$ and $\tau_3$ are 0.23, 0.82 and 0.31, respectively. The significant positive correlation between the oscillations in $\tau_2$ and the conductance oscillations indicates that the dominant contribution of the second channel to the oscillations is a general feature.

Previously reported calculations for finite Pt chains[11,41] obtained a distribution of channels which is in good agreement with our results. Simulations of channel evolution during chain formation presented in reference 11 capture both the nearly constant transmission of the first channel and the variance in the second channel's transmission. For a linear configuration of the atomic chain, the authors find that three main channels contribute to transport. One channel results from a hybridization of $s$, $p_z$ and $d_{3r^2-z^2}$ orbitals ($m_l = 0$), and two additional channels were found to be dominated by either $d_{yz}$ or $d_{xz}$ orbital contributions ($m_l = \pm 1$), in agreement with the three channels found in our channel analysis. In contrast, for Au atomic chains a very different situation is expected, since transport is governed by a single conduction channel, reflecting the dominant *s* character of states near the Fermi energy[1,3,4,42].

The differences in orbital character between the two metals are manifested in the oscillation amplitude during the formation of Au and Pt chains. The significant variation between the two cases can be seen in the 2D conductance-displacement histograms (Fig. 3a,b, top panels), where the color code represents the number of times that a certain combination of conductance and relative electrode displacement was detected. In the Pt histogram, the conductance values of the last plateau exhibit a wide spread, where 95% of the values are in the range of 1.1-2.5$G_0$ and the conductance oscillations are clearly visible. For Au, the last plateau has a conductance of approx. 1$G_0$ with a much smaller spread (0.75-1.1$G_0$) and the relatively smaller oscillations are not apparent. In this case, the single channel composition is clearly manifested as a sharp cut in the probability to obtain conductance values above 1$G_0$ for elongated chains. The fourfold difference in the dispersion of conductance is in good agreement with chain formation simulations, where transitions between zigzag and linear configurations resulted in conductance values in the range of 0.9-2.1$G_0$ for Pt[33] and 0.8-1.0$G_0$ for Au[35].

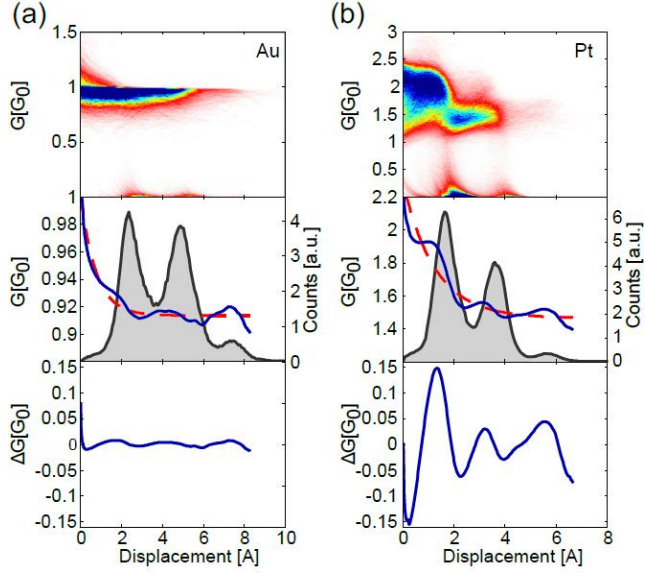

**Figure 3.** Analysis of the oscillation amplitude for Au (a) and Pt (b) atomic chains. Top panels display 2D conductance-displacement histograms for the last plateau, aligned by the first conductance drop below $1.2G_0$ (Au) and $2.5G_0$ (Pt). Middle panels show length histograms (gray area), average traces (blue line) and exponential fits ($y=ae^{-bx}+c$) to the average trace (dashed red). Bottom panels show the differences between the average traces and their exponential fits. The histograms are constructed from the same data sets used in Fig. 1.

To examine the differences in oscillation amplitude more closely, we retrieve the clean oscillation signature by subtracting a baseline from the average trace (Fig. 3a,b, middle panels). The resulting traces (bottom panels) show that the amplitude of the Pt oscillations is about an order of magnitude larger than that of Au. We can therefore conclude that the conductance of Pt chains is substantially more sensitive to the transitions between zigzag and linear configurations compared to Au chains.

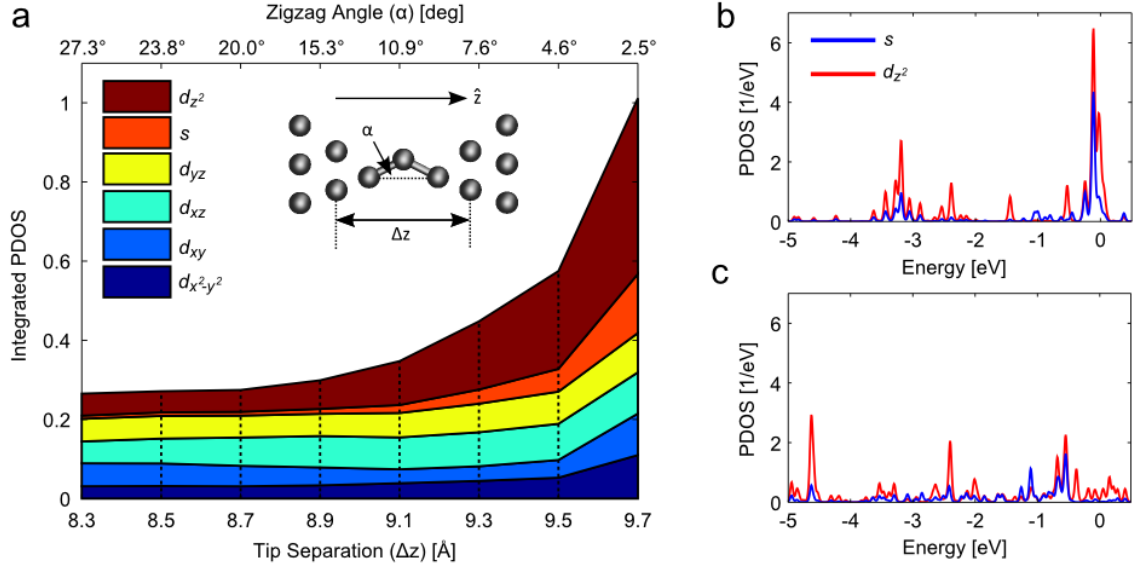

**Figure 4.** (a) Orbital contributions to the localized states at the Fermi energy during simulated stretching of an atomic Pt chain. The number of states shown is obtained for the central atom in an energy window of 100meV centered at the Fermi energy. Inset shows the calculated atomic configuration for Δz=8.3Å. The tip separation and the zigzag angle are defined as indicated in the inset. (b,c) Density of states projected on the central atom for *s* (blue) and $d_{z^2}$ (red) orbital contributions, calculated for linear (Δz=9.7Å; b) and zigzag (Δz=8.3Å; c) chain configurations.

In order to understand the origins for the intense conductance oscillations observed in Pt chains, we performed first principles calculations using density functional theory (DFT)[43]. The calculations were done using VASP[44] code, applying the Perdew–Burke–Ernzerhof (PBE) exchange-correlation functional[45]. The model system consisted of two atomic tips oriented along the <111> direction, bridged by a suspended atom (Fig. 4, inset). The stretching of the atomic chain was simulated by fixing the positions of the tip atoms for a range of tip separations, with the positions of the bridge atom and the two apex atoms allowed to relax (see Supporting Information[38]).

As the atomic contact is stretched, we find that the chain undergoes a transition from a zigzag to a more linear arrangement. At a small tip separation (Δz=8.3Å), an angle of 27° with respect to the chain axis is formed between the three chain atoms. The chain gradually straightens to a nearly linear configuration (angle of 3°) when the tip separation is increased by 1.4Å, in good agreement with previous calculations[33].

The main bottleneck for transport is expected to be located at the central atom, due to the small cross-section of the charge density at the single-atom constriction. Thus, we have obtained the projected density of states (PDOS) on the bridge atom for each of the optimized geometries. We focus on the states near the Fermi energy that are available for transport. Figure 4a shows the PDOS integrated over an energy window of 100meV around the Fermi energy, as a function of the tip separation, partitioned by the angular momentum component of each state (in the frame of reference of the entire system). As can be seen, the overall density of states around the Fermi level increases significantly as the chain is stretched to a more linear configuration. Interestingly, the main contribution to this change comes from states with $m_l=0$, which can be identified as the $s$ and $d_{z^2}$ orbitals of the central atom, the latter being particularly dominant. Conversely, the contributions from orbitals with larger angular momentum ($m_l \geq 1$) are found to be less sensitive to the change in geometry. This behavior is in good agreement with experimental findings. The significant increase in the PDOS near the Fermi level for the linear configuration is consistent with the observed higher conductance. Furthermore, the large variations in the transmission of the secondary channel can be correlated to the availability of $s$ and $d_{z^2}$ states at the Fermi energy, while the invariance in the contribution from $d$ orbitals with $m_l \geq 1$ are in line with the nearly constant transmission of the dominant channel. We have repeated the calculations for a chain of two atoms between the apices, and obtained the same qualitative results (see Supporting Information[38]).

These observations can be understood within the intuitive chemical picture of hybridization between atomic orbitals of the center and apex atoms. Due to the higher symmetry in the linear configuration, the hybridization between the $s$, $d_{z^2}$ orbitals of the central atom and the rest of the $d$ orbitals on the apex-atom is unfavorable, resulting in a relatively energy-localized PDOS (Fig. 4b). In the zigzag configuration, symmetry breaking occurs, allowing considerable hybridization. Consequently, the PDOS of $s$, $d_{z^2}$ undergoes a spreading and shifting to lower energies, reducing the number of states at the Fermi energy (Fig. 4c). Remarkably, the better hybridization in the zigzag configuration results in a counter-intuitive decrease in the number of states available for transport.

These findings can be well visualized using the real space distribution of the charge density at the Fermi level (Fig. 5). For the linear chain (top), the significant contribution of the $d_{z^2}$ orbital is manifested in the shape of the charge distribution around the central atom. The shape of the charge distribution changes to an apparent $d_{xz}$ character in the zigzag case (bottom), indicating a significant shift of orbital character due to the different hybridization with the tip atoms.

Although the conductance oscillations are also observed for Au, the near-Fermi electronic structure for the two systems is different. The density of states of Au around the Fermi energy is dominated by an *s* character, while the *d* orbitals contribute at lower energies and do not readily hybridize with the *s* orbital[46]. The density of states near the Fermi energy is therefore less affected by changes in the local geometry, explaining the order of magnitude decrease in the oscillation amplitude observed experimentally for Au chains.

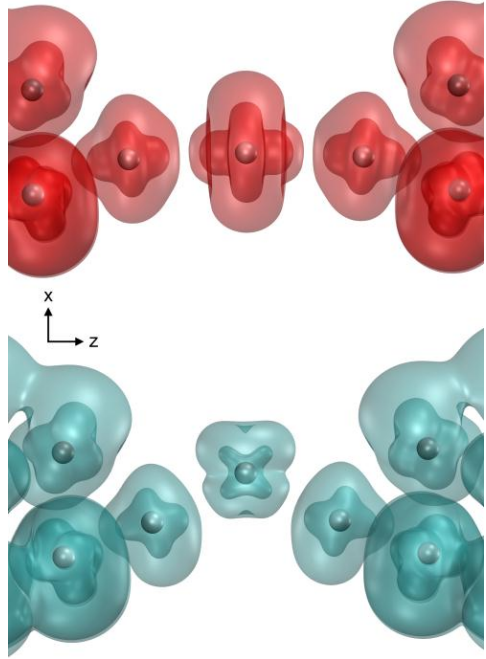

**Figure 5.** Spatial distribution of the charge density near the Fermi level for a linear (Δz=9.7Å; top) and zigzag (Δz=8.3Å; bottom) chain. The charge distribution was calculated for a window of 100meV centered at the Fermi energy, presented with iso-surface values of 90% and 99% of the charge as darker and lighter shaded regions, respectively.

In conclusion, we have found that Au and Pt atomic chains exhibit conductance oscillations during stretching due to repeated transitions between zigzag and linear chain structures. From our combined experimental and computational analysis, we infer that the oscillations result from variations in orbital overlaps, which are substantially more pronounced when symmetry and directionality considerations arising from the $d$ orbitals apply. This conclusion is consistent with the significantly smaller oscillation amplitude measured for Au chains, where electron transport takes place through a single channel dominated by a spherically symmetric $s$ orbital. The presented principle of conductance sensitivity to the spatial symmetry of orbitals can be useful for orbital-oriented engineering of atomic scale conductors. In particular, directional orbitals can be used to promote conductance manipulations by orientation changes, and the use of conductors with spatially isotropic orbitals at the relevant energies can reduce sensitivity to structural variations.

## ASSOCIATED CONTENT

**Supporting Information**. Details regarding the experimental setup, shot noise measurements, data analysis and theoretical calculations, and additional experimental results showing the observation of parity oscillations in superposition with one-atom oscillations. This material is available free of charge via the Internet at http://pubs.acs.org.

## AUTHOR INFORMATION

**Corresponding Author**

E-mail: oren.tal@weizmann.ac.il

**Notes**

The authors declare no competing financial interest.

## ACKNOWLODGEMENTS

The authors are thankful to J. M. van Ruitenbeek and F. Pauly for helpful discussions. O.T thanks the Harold Perlman Family for their support and acknowledges funding by the Israel Science Foundation, the German-Israeli Foundation and the Minerva Foundation. L. K. Acknowledges support from the Israel Science Foundation and the Lise Meitner Center for Computational Chemistry.

# Probing the Orbital Origin of Conductance Oscillations in Atomic Chains

**Supporting Information**

*Ran Vardimon[1], Tamar Yelin[1], Marina Klionsky[1],*
*Soumyajit Sarkar[2], Ariel Biller[2], Leeor Kronik[2] and Oren Tal[1]*

[1] Department of Chemical Physics, Weizmann Institute of Science, Rehovot 76100, Israel
[2] Department of Materials and Interfaces, Weizmann Institute of Science, Rehovot 76100, Israel

**Table of contents**

## S1. Experimental Setup

Atomic chains are formed using a mechanical controllable break junction setup operated in cryogenic temperature conditions (4.2K). The sample consists of an Au or Pt wire (0.1mm diameter, 99.99% purity), which is notched at its center and fixed to a flexible phosphor bronze substrate covered by an insulating Kapton film (Fig. 1a). Using a piezoelectric element, the sample is bent until the wire breaks in its weak spot, forming two atomically sharp tips. A triangular voltage waveform is applied on the piezoelectric element in order to repeatedly break and reform the atomic contact. A constant bias voltage (10-100mV given by a NI PCI 4461) is applied across the contact and the current is amplified by an IV preamplifier (SR570) and recorded during the contact evolution. The pulling speed for the presented data is 600nm/s for the Pt traces and 300nm/s for Au.

Noise measurements were performed by amplifying the voltage signal with two sets of low noise voltage amplifiers (NF LI-75a followed by Signal recovery 5184) and then calculating the cross-correlation between the two recorded signals (Fig. 2a) using a SR785 dynamic signal analyzer. Clean bias current is supplied by a 4461 DAQ card connected to sample by two 500kΩ resistors. The relatively noisy components of the conductance measurement circuit are disconnected from the sample when performing noise measurements. At each bias step, the measured spectrum (Fig. 2b) is obtained by averaging 5,000 recorded spectra. The averaging is performed in order to eliminate non-correlated voltage noise originating from the amplifiers.

## S2. Shot noise measurements and data analysis

Shot noise measurements are performed by recording sequences of electronic noise spectra during elongation of atomic Pt contacts. For each contact configuration, a set of noise measurements is obtained as function of applied bias voltage (Fig. S1a). The noise spectra were measured at a frequency range between 0-100kHz. The noise generated by the sample is frequency independent in the measured range, however, at relatively high frequencies, the measured noise is reduced due to the low pass transfer function of the electronic circuit. The original signal is recovered by correcting the spectra according to a fitted RC transfer function (Fig. S1b). The average current noise $S_I(V)$ is then calculated from a selected frequency window that is free of spurious spikes. Figure S1c shows the dependence of $S_I(V)$ on applied bias voltage.

The current noise generated by a ballistic conductor is given by[1]:

$$(1)\ S_I(V) = 4k_B T G_0 \sum_{n=1}^{N} \tau_n^2 + 2eV \coth\left(\frac{eV}{2k_B T}\right) GF$$

where $\tau_n$ are the transmissions probabilities of N conduction channels, $F = \sum_{n=1}^{N} \tau_n(1 - \tau_n) / \sum_{n=1}^{N} \tau_n$ is the Fano factor, and $e, V, k_B, G_0, T, G$ are the electron charge, applied bias voltage, Boltzmann constant, conductance quantum, temperature and the conductance of the sample, respectively. At zero bias voltage, the expression reduces to the Johnson-Nyquist expression for thermal noise $S_I = 4kTG$, while at high bias voltage ($eV \gg kT$) the noise has a linear dependence on bias $S_I = 2eVGF$. In order to extract the Fano factor, we follow the procedure introduced by Kumar et al.[2]. Two parameters are used to obtain a linear expression from which the Fano factor can be determined:

$$(2)\ Y(V) = \frac{S_I(V) - S_I(0)}{S_I(0)}$$

$$(3)\ X(V) = \frac{eV}{2k_B T} \coth\left(\frac{eV}{2k_B T}\right)$$

Using the reduced parameters, equation (1) can now be written as a linear relation:

$$(4)\ Y(V) = [X(V) - 1]F$$

The Fano factor is obtained from the slope of a linear fit to eq. (4). An example fit is shown in Fig. S1d. The errors in $S_I$ are determined from the standard deviation of the noise within the selected frequency window. The final error in the Fano factor is determined from the accuracy of the linear fit.

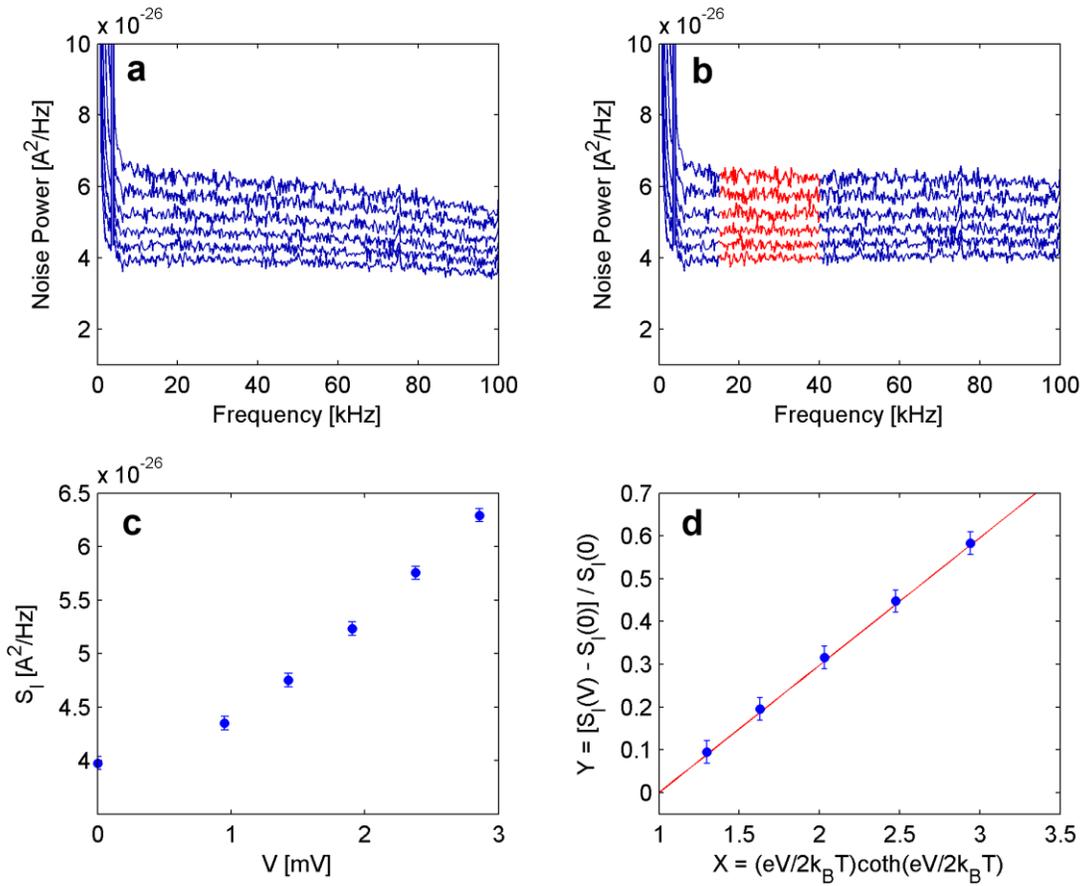

**Figure S1. a**, Electronic current noise spectra measured on a single-atom Pt contact ($G=1.71G_0$) as function of bias voltage. **b**, Spectra corrected for the RC transfer function of the electronic circuit (blue). The red colored section indicates the frequency window from which the average noise is calculated. **c**, Average noise as a function of bias voltage. **d**, The data plotted using the reduced parameters, X and Y (blue points) and a linear fit (red), which gives a Fano factor of $F = 0.30\pm0.01$.

Each series of noise measurements performed on a specific realization of an atomic contact is preceded and followed by measuring the differential conductance (dI/dV) of the contact as function of bias voltage (Fig. S2). The measurements are performed using a lock-in technique with a reference signal frequency of ~3kHz. The purpose of these measurements is twofold. First, they are used to verify the stability of the contact during noise measurements. Similar dI/dV spectra before and after the noise measurements (i.e. as in Fig. S2) indicate that the atomic configuration of the contact did not change during the time of the measurements. Second, the conductance of the contact is determined from the average conductance measured within the range of bias voltages used for shot noise measurements (shown as dashed black lines in Fig. S2). The error in the conductance $\Delta G$ is determined as half the difference between the maximum and minimum conductance values measured in the specified bias range, for the data recorded both before and after the noise measurements. Measurements with an error of $\Delta G > 0.03$ are discarded, ensuring that the conditions of eq. (1), i.e. relatively unvarying transmission probabilities, are fulfilled.

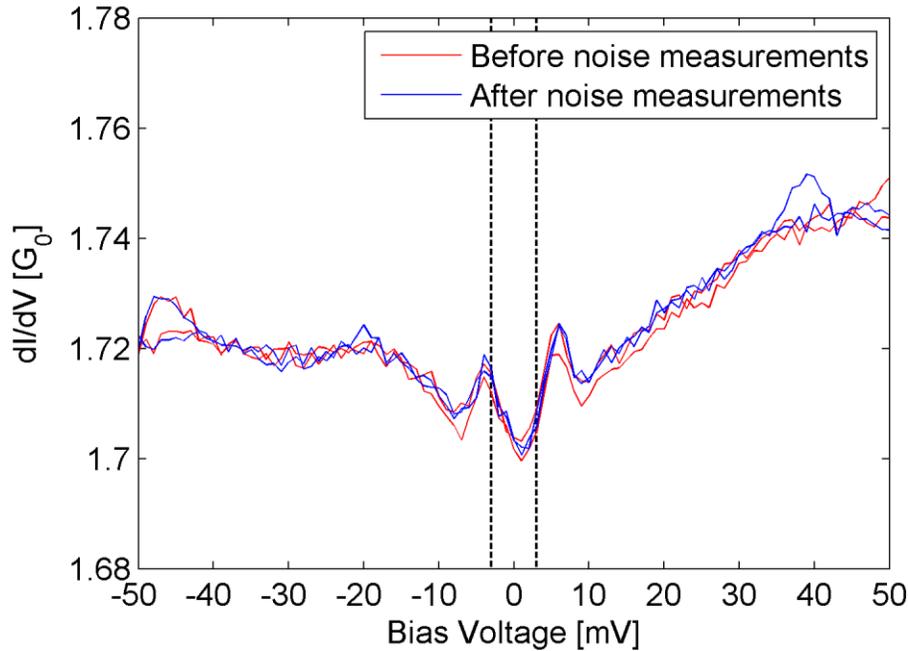

**Figure S2**. Differential conductance (dI/dV) as function of bias voltage measured on a single atom Pt contact. Red/blue curves show the different conductance measured before/after the series of noise measurements. Dashed black curves indicate the region from which the average conductance is calculated.

## S3. Observation of parity oscillations

Smit *et al.* have previously reported on parity oscillations in the conductance of Au and Pt atomic chains with a two-atom period[3] while oscillations with a one-atom period were not reported. Working with different contacts, we find that the manifestation of these two different effects in the average trace can vary between measurements. Using the same sample, it is possible to observe oscillations with one-atom or two-atom periodicities, and in some cases both types of oscillations can even coexist. Figure S3 shows a measured Pt data set in which both effects take place. The oscillation pattern in the average trace (top panel) consists of a superposition of both one-atom period oscillations and oscillations with a double period, as seen more clearly after subtracting the fitted baseline (bottom panel). The data sets in Fig. 3b (in the main text) and in Fig. S3 were obtained using the same sample, however between the measurements of the different data sets the contact geometry was reshaped by strongly reforming the two tips of the junction towards each other by more than 500Å. Although conductance oscillations with one-atom period are more common than parity oscillations, the later can occasionally be detected by repeating the mentioned practice.

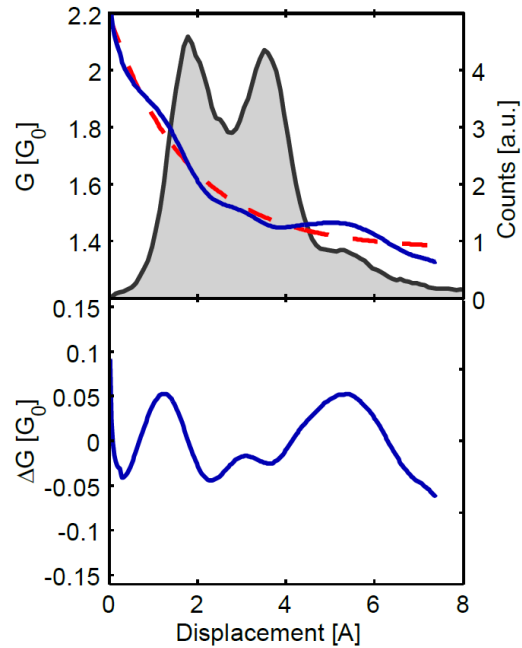

**Figure S3**. Average conductance of Pt atomic contacts showing a superposition of one-atom and two-atom oscillations. Top panel shows length histogram (grey area), average trace (blue line) and an exponential fit to the average trace (dashed red). Bottom panels show the differences between the average traces and their exponential fits. The data consists of 10,000 traces measured on the same sample used in Fig. 3(b).

The average trace shows a superposition of both types of oscillations, the zigzag oscillations typically having the larger amplitude and thus more visible (as in Fig. 3 of the main article). The geometry of the contact and the extent by which similar configurations are reached in different pulling sequences determine by how much are the features in the average trace and length histogram averaged out. Zigzag oscillations are expected to be more sensitive to such averaging since the oscillation wavelength is twice as shorter than for even-odd oscillations[4]. Therefore, in cases that significant averaging occurs (e.g. due to more significant phase shifts between individual traces), the appearance of zigzag oscillations in the average trace is more effectively suppressed and the even-odd oscillations are consequentially more visible.

## S4. Computational details

We carried out first-principles calculations within the framework of density functional theory (DFT)[5] by employing the Perdew–Burke–Ernzerhof (PBE) exchange-correlation functional[6] in a projector augmented plane-wave (PAW)[7] formulation as implemented within the VASP code[8]. The plane-wave cutoff energy was chosen as 550eV, with a Fermi-Dirac smearing of the occupations in an energy equivalent to 15meV. A single k point was used, in a simple orthogonal cell of dimensions 27.1x27.1x40.6Å. In order to deal with the finite extent of our model systems within the adopted periodic scheme of calculations, the dimensions of the orthogonal cell were chosen such that the edges of two neighboring images were kept separated by a vacuum of about 13–16Å. This makes sure that the interactions between the system images are negligible for any practical purpose.

The stretching of the junction was simulated by varying the tip separation, while the unit-cell dimensions were kept constant. The tip separation is defined as the normal distance between the planes containing the three atoms which are located immediately behind the apex atom of each electrode. The positions of the apex and wire atoms were optimized without symmetry constraints such that each component of force on every atom does not exceed $5meV\text{Å}^{-1}$. In this study, we have considered the range of 8.3-9.7Å, in which the 3 atom chain was found to be energetically stable. For smaller tip separations, the chain returned to a linear configuration, however the Pt-Pt bond length between the chain atoms was found to be significantly lower than the equilibrium bond length for bulk Pt and the total energy increased substantially. For tip separations above 9.7 Å, the Pt-Pt interaction between atom and apex is so weak, that a slight perturbation to the position of the central atom pushes the atom towards a configuration where it is only bound to one of the tips, suggesting the breaking of the junction in experiment (or the pulling of an additional atom into the chain).

We have tested the effect of varying the number of planes (n) constituting the atomic tip. Structures with n > 3 were found to produce similar results, and thus we have fixed n=5. Note that due to the tetrahedral symmetry of each tip, there is a dipole moment in the system unless an inversion center is enforced by creating a relative angle between tips. However, in this case the correspondence with the cell's spherical harmonics will be less than optimal, and thus a system with a σ plane was chosen instead. It was checked and verified that for such a system where tips are mirror images of each other, the dipole is adequately small at the central atom position. In addition, we have checked the system with inversion symmetry, to see if there is some effect on forces and the total DOS and found that this does not have any notable influence in the distances and angles studied in this work.

Calculating the projected density of states over spherical harmonics is a standard operation in VASP which provides information regarding the local density of states at a specified region in the unit cell, typically around selected atoms. Since the states are projected over spherical harmonics in the system's coordinates, the orientation of the chain and tips were chosen in advance such that the chain axis is along the $\hat{z}$ direction, and such that the zigzag angle is within, or very near to, the xz plane.

## S5. Effect of structural configuration on the orbital contributions at the Fermi level

To study the effects of the chain geometry on the electronic structure, we have calculated the projected density of states (PDOS) on the central chain atom, which constitutes the bottleneck for electron transport due to its small cross-section. We focus on the states in an energy window of 100meV centered at the Fermi energy, since only these states are relevant for the conducting charge carriers, according to the maximal bias voltage of 100mV used in the experiment. As discussed in the main text, the increase in the PDOS for the linear chain configuration originates from the contribution of states with zero angular momentum, namely $s$ and $d_{z^2}$. To better understand this effect, we obtained the PDOS of these orbitals, as presented in Fig. S4. As one can see, the PDOS of $s$ (a) and $d_{z^2}$ (b) states near the Fermi energy undergoes a significant increase as the chain is stretched to a more linear configuration. For the initial zigzag configuration, both states contribute very little to the PDOS. However, when the chain is brought to a more linear configuration, a resonance is gradually built up and shifted towards the Fermi energy.

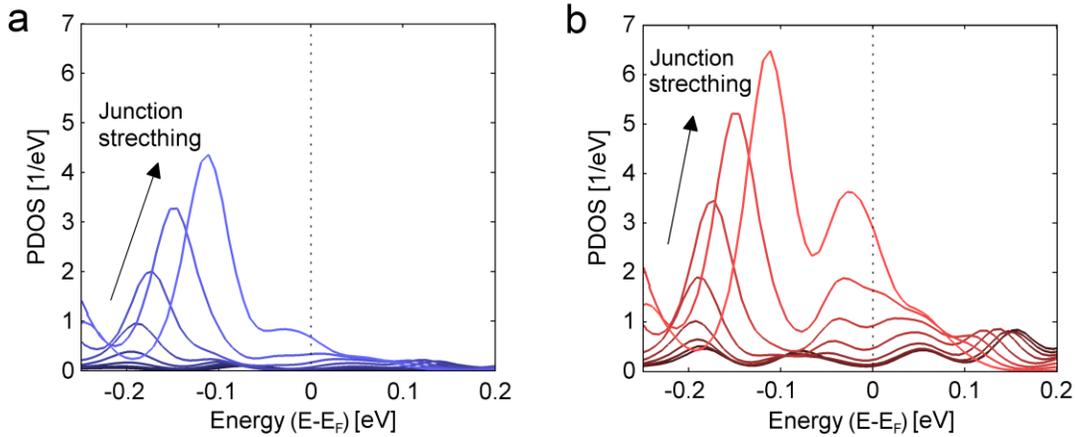

**Figure S4.** Density of states projected on the central atom for $s$ (a) and $d_{z^2}$ (b) orbital contributions, calculated for different tip separations between 8.3 Å to 9.7Å.

The effect of the zigzag angle on the PDOS near the Fermi energy was investigated for both the central bridge atom and the tip apex atom. For each case, we have calculated the integrated PDOS in an energy window of 100meV around the Fermi level, partitioned by angular moment contributions (in the reference frame of the entire system). The calculations were repeated for a four atom chain in order to check whether the effect is consistent. The results are shown in Fig. S5 for the 3 atom chain (a,b) and 4 atom chain (c,d). As one can see, the results for two cases are qualitatively similar. While in both cases the integrated PDOS for the bridge atoms increases with the tip separation, for the apex atoms the density of states is insensitive to junction elongation. This insensitivity could be explained by the fact that the apex atom is bonded to three neighbors on one side, stabilizing its orbital structure.

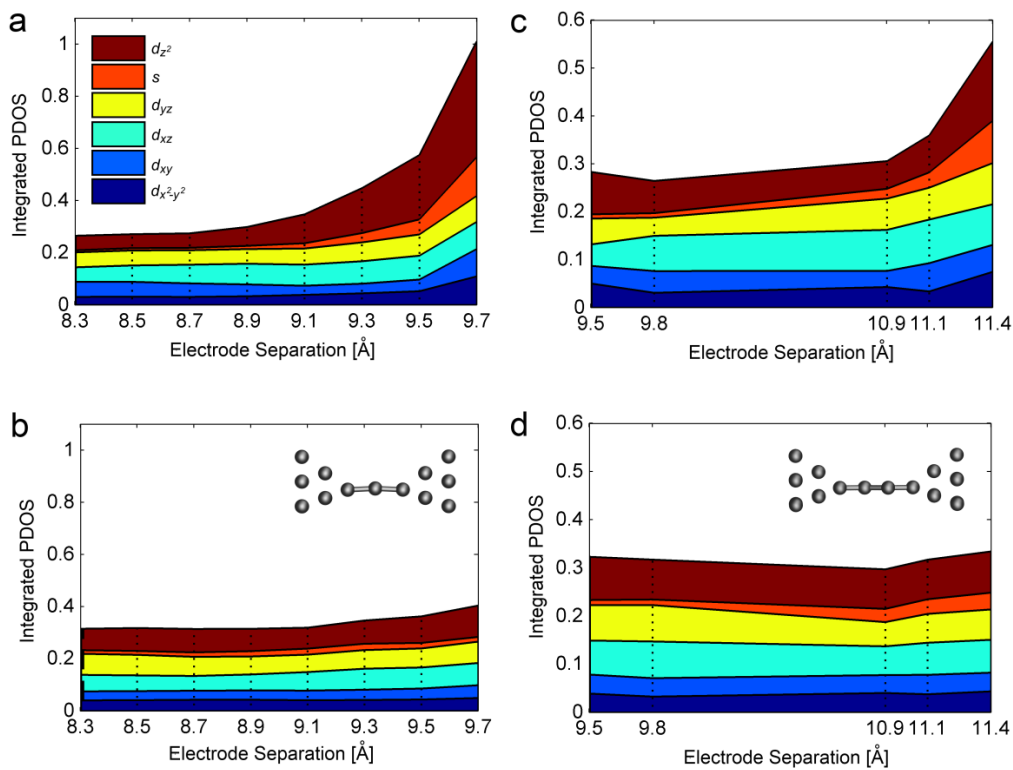

**Figure S5.** Orbital contributions to the localized states at the Fermi energy during simulated stretching of an atomic Pt chain consisting of three (a,b) and four (c,d) atoms. The number of states shown is obtained for a bridge atom (a,c) and a apex tip atom (b,d) in an energy window of 100meV centered at the Fermi energy.

Note that for the four atom zigzag chain, the stable geometry cannot be rotated so that all the atoms are offset from the z axis on a single plane. Therefore, the system's spherical harmonics can be less accurately applied to identify the atomic angular content of orbitals at the chain atoms. Nevertheless, the integrated PDOS still exhibits the enhancement effect upon transition towards a more linear configuration. Figure S6 shows the real space distribution of charge density for a 4-atom chain in zigzag configuration. Similar to the case of the 3-atom chain, the local charge distribution on the bridge atoms shows an apparent $d_{xz}$ character, while the absence of the $d_{z^2}$ contribution again results from the hybridization with the remaining $d$ orbitals.

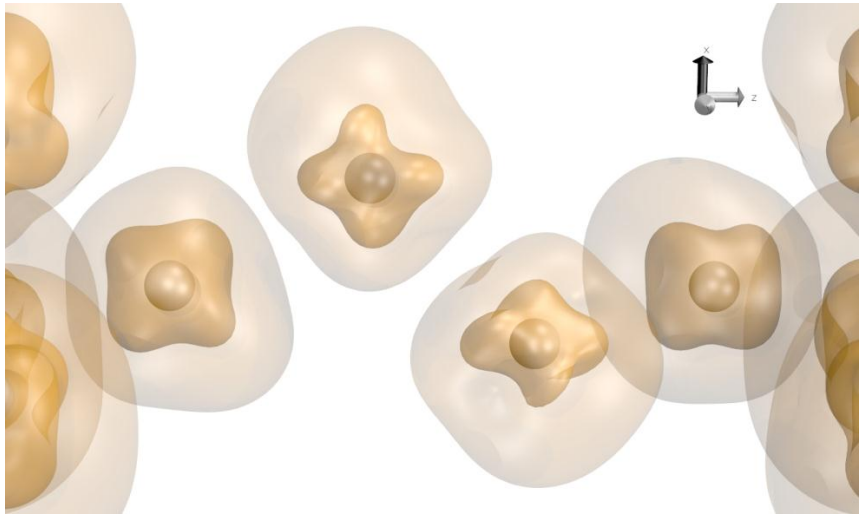

**Figure S6.** Spatial distribution of the charge density near the Fermi level for a zigzag ($\Delta z=9.5$Å) chain. The charge distribution was calculated for a window of 100mV centered at the Fermi energy, presented with iso-surface values of 90% and 99% of the charge as darker and lighter shaded regions, respectively.